\documentclass[%
 reprint,
 amsmath,amssymb,
 aps,
 onecolumn,
 12pt,
]{revtex4-2}

\usepackage{graphicx}
\usepackage{subcaption}
\usepackage{float}
\usepackage{dcolumn}
\usepackage[export]{adjustbox}

\usepackage{bm}

\begin{document}

\preprint{APS/123-QED}

\title{Impacts of packed bed polydispersity and deformation on fine particle transport}


\author{Dhairya R. Vyas$^1$}
\author{Song Gao$^1$}
\author{Paul B. Umbanhowar$^1$}
\author{Julio M. Ottino$^{1,2,3}$}
\author{Richard M. Lueptow$^{1,2,3}$}
\affiliation{%
\\$^1$Department of Mechanical Engineering, Northwestern University, Evanston, Illinois 60208, USA.\\
$^2$Department of Chemical and Biological Engineering, Northwestern University, Evanston, Illinois 60208, USA.\\
$^3$Northwestern Institute on Complex Systems (NICO), Northwestern University, Evanston, Illinois 60208, USA.
}%


\begin{abstract}
        Static granular packings play a central role in numerous industrial applications and natural settings. In these situations, fluid or fine particle flow through a bed of static particles is heavily influenced by the narrowest passage connecting the pores of the packing, commonly referred to as pore throats, or constrictions. Existing studies predominantly assume monodisperse rigid particles, but this is an oversimplification of the problem. In this work, we illustrate the connection between pore throat size, polydispersity, and particle deformation in a packed bed of spherical particles. Simple analytical expressions are provided to link these properties of the packing, followed by examples from Discrete Element Method (DEM) simulations of fine particle percolation demonstrating the impact of polydispersity and particle deformation. Our intent is to emphasize the substantial impact of polydispersity and particle deformation on constriction size, underscoring the importance of accounting for these effects in particle transport in granular packings.
\end{abstract}

\maketitle

\section{Introduction}

The packing of particles holds significant relevance across various fields, including fluid and particle flow in porous media for filtration \cite{roozbahaniMechanicalTrappingFine2014, xiaoGranularBedFilter2013a,yuFiltrationPerformanceGranular2020a}, soil mechanics \cite{panayiotopoulosPackingSandsReview1989}, sedimentation of particles \cite{jerkinsOnsetMechanicalStability2008, houssaisOnsetSedimentTransport2015}, and geological phenomena \cite{pufahlSedimentaryIgneousPhosphate2017, itohGeologicalImplicationGrainsize2018,changJammingDensityVolumePotential2022}. A comprehensive understanding of the properties of static packed granular beds, including packing density, permeability, and pore volume, provides valuable insights into these fields.
%
%
A crucial aspect of particle packings is the characterization of pore throat size or constrictions. Pore throats are the narrowest constrictions within the interconnected pore network in a bed of particles, where the flow of fluids or particles is most restricted. These constrictions significantly influence the flow behavior and transport properties in the packed medium. Understanding the distribution and dimensions of pore throats is essential in various applications, such as enhanced oil recovery \cite{gaoQuantitativeDeterminationPore2016}, where the movement of oil through reservoir rocks is influenced by the size of the narrowest passages. Similarly, in groundwater studies \cite{nsirPorethroatModelBased2010,georgeHydrodynamicImplicationsAquifer2017}, pore throat size can control the flow rate and retention of contaminants in an aquifer. Pore throat size also affects the transport of fine particles in industrial or pharmaceutical solids handling where the fine particles are small enough to pass through the voids between stationary large particles. This leads to degraded product quality, fouled equipment, health risks like inhalation of fine particles, and safety hazards such as dust explosions \cite{bemroseReviewAttritionAttrition1987,PowdersBulkSolids2007}.  Therefore, accurate characterization of pore throat size distribution is important. 

\begin{figure}[H]
    \centering
    \begin{subfigure}[b]{0.27\textwidth}
        \includegraphics[width=\linewidth]{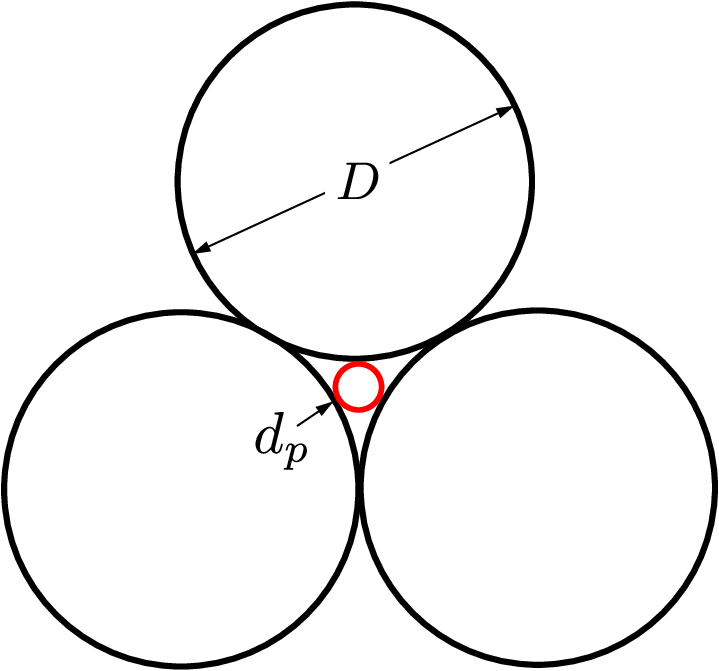}
        \caption{}
    \end{subfigure}
    \hspace{0.05\textwidth}
    \begin{subfigure}[b]{0.23\textwidth}
        \includegraphics[width=\linewidth]{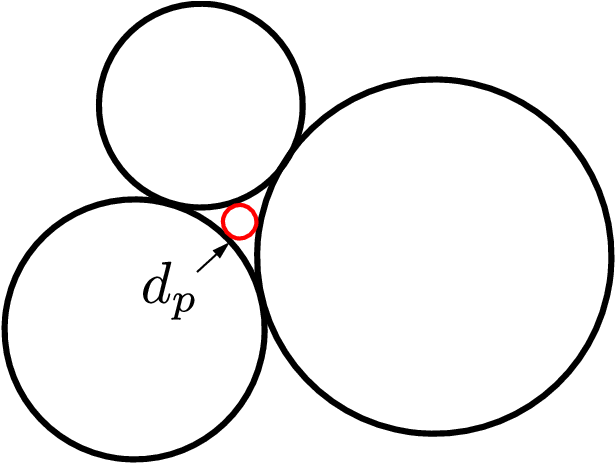}
        \caption{}
    \end{subfigure}
    \hspace{0.05\textwidth}
    \begin{subfigure}[b]{0.27\textwidth}
        \includegraphics[width=\linewidth]{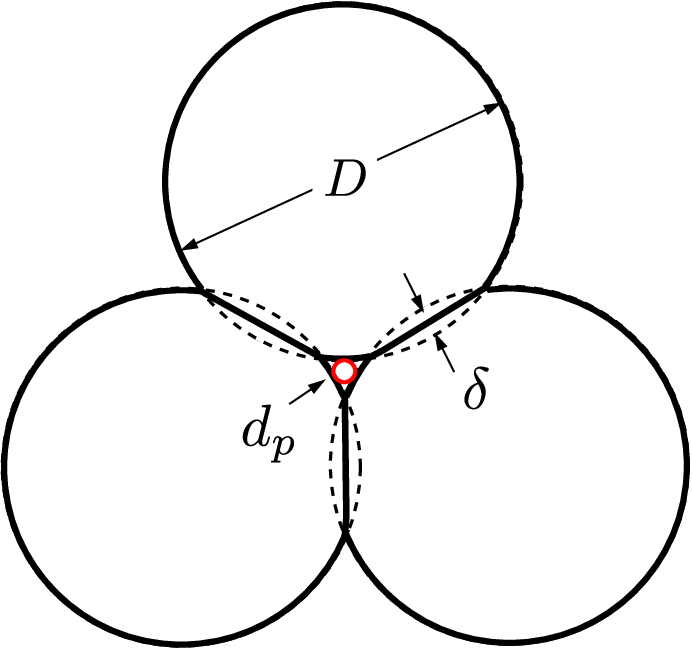}
        \caption{}
    \end{subfigure}
    \caption{Pore throat circle (red) formed by (a) three identical rigid particles in tangential contact; (b) a large particle with two slightly smaller particles that are in mutual tangential contact; and (c) three identical deformed particles in contact, with deformation represented as overlap $\delta$.}
    \label{fig:sch}
\end{figure}

The idealized problem for the packing of spherical particles is most simply framed in terms of three contacting circles, as shown in Fig.~\ref{fig:sch}(a), the study of which can be traced back to the ancient Greeks, with Apollonius of Perga (c. 240 - c. 190 BC) being one of the first mathematicians to study it. 
The “circles of Apollonius” is a series of problems that includes three mutually tangent circles that define a smaller inscribed circle that is tangent to all three circles \cite{matthewsStaticsAnalyticalGeometry2019}. This smaller inscribed circle represents the most restrictive passage or pore throat, that is generated when three large spherical particles are in contact (see Fig.~\ref{fig:sch}(a)). 
For our specific interest in solids handling applications, the smallest pore throat determines the size of the largest fine particle that can freely flow between the interstices, under the effect of an external force like gravity, a flow behavior commonly referred to as free sifting. Hence, we use this application as the context for our study.  However, many other applications depend on the passage or blockage of fine particles or fluids through pore throats in a bed of spherical particles for which these results are useful.

Central to understanding free sifting in  granular beds composed of spheres is identifying the particle size ratio, above which fine particles freely pass through the smallest possible voids between static large particles. Dodds \cite{doddsPorosityContactPoints1980a} considered this problem in detail as related to the porosity of random sphere packings. The size ratio of the large sphere diameter ($D$) to the diameter of the largest fine particle that can pass through the smallest pore throat is $R_{t0} = D/d_p = \sqrt{3}/(2 - \sqrt{3}) \approx 6.464$, where $d_p$ is the pore throat diameter and the zero subscript for $R_{t0}$ indicates that this is an idealized value for the trapping threshold, $R_t$ \cite{ippolitoDiffusionSingleParticle2000a,lomineDispersionParticlesSpontaneous2009b}, based purely on the geometry shown in Fig.~\ref{fig:sch}(a). Numerical and experimental studies conducted after the work of Dodds \cite{doddsPorosityContactPoints1980a} show that the actual trapping threshold for a static randomly packed large particle bed is $R_t = 6.67$, which is slightly larger than $R_{t0}$ due to the occasional formation of jammed arches of fine particles between large particles \cite{roozbahaniMechanicalTrappingFine2014,kerimovMechanicalTrappingParticles2018}. Similarly, for deep bed filtration, where multiple fine particles in suspension are captured by a granular bed, the particle size ratio at the filtration threshold is observed to be about 6.62 \cite{ghidagliaTransitionParticleCapture1996}, which again is slightly larger than $R_{t0}$. In spite of these measured values for $R_t$ differing from the theoretical value, $R_{t0}$, the trapping threshold size ratio is an important parameter that can be used to characterize the sizes of fine particles that will either pass through or be trapped in granular packings.

Here, we analyze two variations of the trapping threshold that are relevant to practical aspects of the fine particle percolation problem \cite{gaoPercolationFineParticle2023a} as well as many other problems related to packed beds of particles.  Both variations emerge from geometric factors but appear to not have been previously addressed in the literature.  If so, this is surprising, given that the history of circle packings goes back to ancient times \cite{NewApproachesCircle2007}.

In sphere packings of real and model materials, it is common that particles vary in size.  This can come about from the natural size distributions of the particles or it may be desirable to vary the size of the bed particles to avoid an inadvertent crystal-like packing when a random packing is preferred, such case might occur in Discrete Element Method (DEM) simulations of packed beds of particles \cite{gaoPercolationFineParticle2023a}. The variation in particle size necessary to avoid crystal-like packing is usually relatively small. Variations of about $\pm 10$\% around the mean value typically result in random packings \cite{campoDynamicalCoexistenceModerately2020, bandeltriessAssessingWallEffects2022,haoRheologyShearedPolyhedral2023}. However, the impact of polydispersity on the size of pore throats can be significant (see Fig.~\ref{fig:sch}(b)). Therefore, the first issue we explore relates to the influence of size-polydispersity of the large bed particles on the trapping threshold.

The second issue we consider focuses on the influence of large bed particle deformation on the trapping threshold.
The standard theoretical value for the geometrical trapping threshold of $R_{t0}=6.464$ is based on the assumption that the spheres are rigid and in point contact with their neighboring spheres.
However, for all materials, some level of deformation occurs when particles make contact, as occurs in a diverse range of applications including pharmaceutical \cite{sunRelationshipTensileStrength2018} and biomedical applications \cite{dalyHydrogelMicroparticlesBiomedical2020}, nuclear reactors \cite{rycroftGranularFlowPebblebed2013}, ion exchange beds \cite{tiihonenElasticityIonexchangeResin2001}. In these and other scenarios the Young's moduli of the bed particles range from a few kilopascals to several gigapascals: 20 kPa for hydrogels \cite{dalyHydrogelMicroparticlesBiomedical2020} employed in biomedical applications, 0.0027-0.005 GPa for expanded polystyrene \cite{chenStaticDynamicMechanical2015}, 0.001-0.05 GPa for silicone rubber at small strain \cite{PropertiesSiliconeRubber}, 0.2-4 GPa for polymers \cite{amitay-sadovskyEvaluationYoungModulus1998,premalalComparisonMechanicalProperties2002},  48-84 GPa for glass \cite{PropertiesFloatGlass}, 68 GPa for aluminum \cite{nayarMetalsDatabook1997}, and 200 GPa for steel \cite{nayarMetalsDatabook1997}. As a result of both the particle stiffness and different loading scenarios in various applications, the particle deformation can vary substantially, thereby significantly influencing the size of pore throats (see Fig.~\ref{fig:sch}(c)). If sufficiently large, deformation can even result in the complete closing of pore throats. An analogous problem occurs in DEM simulations in which forces between particles are based on the overlap between particles, shown as $\delta$ in Fig.~\ref{fig:sch}(c). 
 
While both of these factors can significantly reduce the pore throat size and influence the geometric trapping threshold, most studies have not considered their influence. For example, in experiments \cite{lomineDispersionParticlesSpontaneous2009b} and DEM simulations \cite{lomineTransitTimeInterparticle2010} on percolation of fine particles, a geometric trapping threshold of $R_t=6.464$ was assumed. However, some degree of polydispersity in particle size is inevitable in experiments and frequently used in simulations. Furthermore, the algorithm used to generate the granular bed (for example see \cite{powellComputersimulatedRandomPacking1980}) can lead to large particle overlap. 
A similar premise of $R_t = 6.464$ was made in another study \cite{liSpontaneousInterparticlePercolation2010} when analyzing the percolation of fine particles without accounting for potential deformation effects. Similar assumptions have been made in other studies \cite{sollerRheologyEvolvingBidisperse2011c,buiModellingSedimentExchange2020,xieCFDDEMModellingMigration2021a} where the effects of polydispersity and deformation (overlap) could be significant but have not been considered.

In this paper, we investigate both of these effects, first from a purely geometric standpoint, and then with respect to practical application in DEM simulations.
Although pore throat characterization holds significant importance in various applications across several fields, our examples are focused on fine particle-free sifting in a dry static packed bed of randomly packed large spherical particles. This simple problem allows us to demonstrate and analyze the influence of polydispersity and deformation on the trapping threshold and gain valuable insights into the mechanics of particle entrapment in a simple and straightforward manner. 

The paper is structured into two primary sections: Section 2 tackles the issue of particle polydispersity, while Section 3 addresses the problem of particle deformation. This enables us to study and analyze these effects separately, leading to the development of distinct analytical formulations for each. These formulations remain applicable even when both particle polydispersity and deformation are simultaneously present, necessitating only minor adjustments.

\section{Effect of size-dispersity on trapping}
\label{sec:poly}

\begin{figure}
    \centering
    \includegraphics[width=0.7\linewidth]{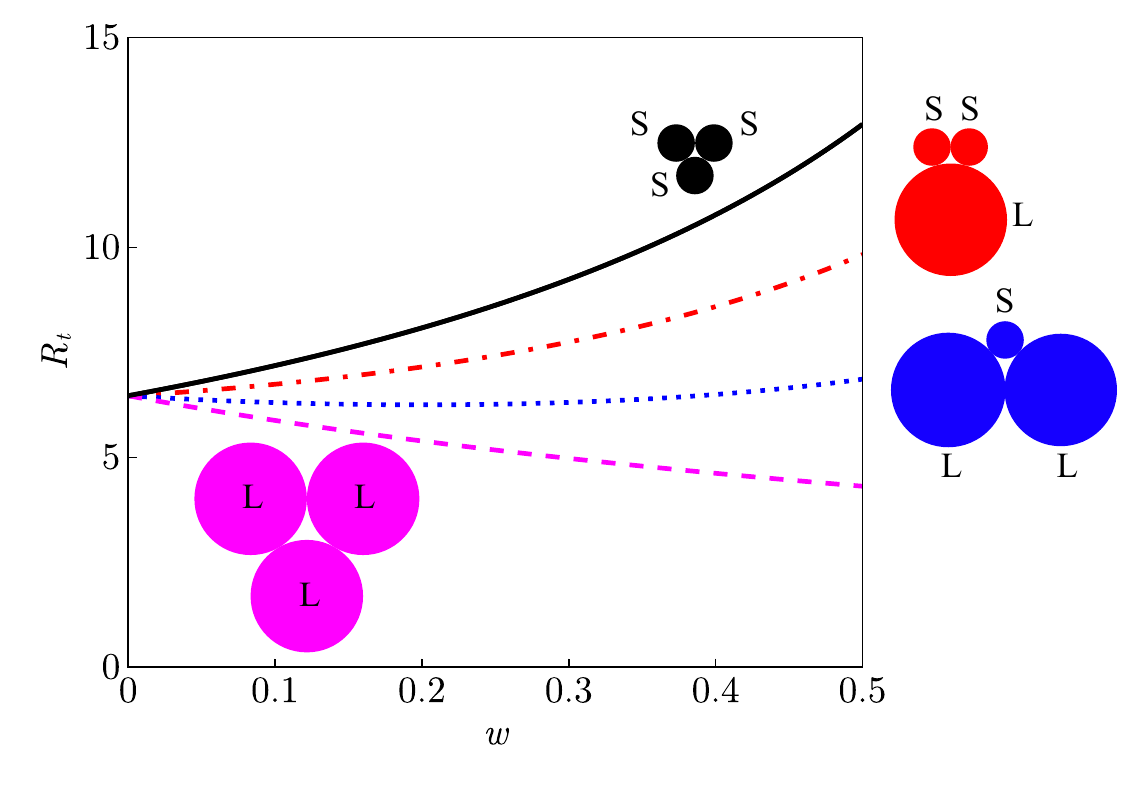}
    \caption{Trapping threshold, $R_t$, vs. bed particle size distribution half-width, $w$, for four extreme situations. SSS: three smallest particles touch (black solid); LLL: three largest particles touch (magenta dashed); SLL: two largest and one smallest particles touch (blue dotted); and SSL: two smallest and one largest particles touch (red dash-dotted).}
    \label{fig:chporefig1}
\end{figure}

When the particles in a static packed bed are size-disperse, i.e., they vary in diameter, the resulting pore throat size distribution formed within the static packing becomes a function of the bed particle size distribution. In particular, the size of the smallest pore throat is crucial because it determines the narrowest passageway through the packed bed, or, equivalently, the largest fine particle that can pass through the bed without being trapped. When the width of the bed particle size distribution is $2w$, bed particle diameter $d$ ranges from $(1-w)D$ to $(1+w)D$, where $D$ is the bed particle diameter corresponding to the midpoint of the particle size distribution. By enumerating the possible combinations of particles with extreme sizes, we can identify four possible scenarios: three smallest particles (SSS), three largest particles (LLL), one smallest and two largest particles (SLL), or one largest and two smallest particles (SSL), see Fig.~\ref{fig:chporefig1}. For each scenario, the three particles are assumed to be tangent in the same plane. In the two-dimensional circle packing (which is equivalent to the three-dimensional packing of three spheres), the pore throat diameter, $d_p$, for these four scenarios, can be found as a function of particle size distribution half-width, $w$. The effective trapping threshold is defined as $R_t=D/d_p$, which can be compared to that for identical bed particles (diameter $D$), $R_{t0}=6.464$. 

The trapping threshold for the SSS case can be found by first noting that $(1-w)D=6.464d_p$, which can be rewritten as
\begin{equation}
R_t=R_{t0}/(1-w).
\label{eq:chporeeq1}
\end{equation}
In this case, $R_t$ increases as $w$ increases, which is shown as the black solid curve shown in Fig.~\ref{fig:chporefig1}. This is reasonable as the increase of $w$ leads to the reduction of the diameter of the smallest bed particles, which necessarily form a smaller pore throat, leading to a larger $R_t$ (equivalent to a smaller fine particle being able to fit through the pore throats of the particle bed).

For comparison, the trapping threshold for the LLL case, for which $(1+w)D=6.464d_p$, is
\begin{equation}
R_t=R_{t0}/(1+w).
\end{equation}
$R_t$ for the LLL case decreases as $w$ increases (magenta dashed curve in Fig.~\ref{fig:chporefig1}). Again, this is reasonable in that larger largest bed particles result in a larger pore throat.

For the SLL condition, we first solve a set of three equations to determine the pore throat radius $r_p$ and the location of its center $(x_p,y_p)$,
\begin{equation}
(x_p-x_i)^2+(y_p-y_i)^2=(r_p+r_i)^2,
\end{equation}
where $i=1,2,3$ corresponds to a touching triad of one smallest and two largest bed spheres. $x_i$ and $y_i$ are the positions of the center of sphere $i$  in the contact plane, while $r_i$ is the radius of sphere $i$. Then the pore throat diameter is obtained using $d_p=2r_p$, and the $R_t=D/d_p$ can be written as
\begin{equation}
R_t=\frac{2w/(1-w)+\sqrt{(3+w)/(1-w)}}{2-\sqrt{(3+w)(1-w)}} .
\end{equation} 

Similarly, $R_t$ as a function of $w$ for the SSL case can be, expressed as
\begin{equation}
R_t=\frac{-2w/(1+w)+\sqrt{(3-w)/(1+w)}}{2-\sqrt{(3-w)(1+w)}}.
\end{equation}
These two cases are plotted as the red dot-dashed curve and the blue dotted curve, respectively,  in Fig.~\ref{fig:chporefig1}. The two cases lie between the cases for equal size bed particles indicating that SSS and LLL determine the largest and the smallest values of $R_t$.


Thus, a slight increase in the bed particle size distribution width can completely change the percolation behavior of a fine particle (with a fixed size) in the packed bed from passing through the packed bed to being trapped in the bed. For example, if the large to fine particle size ratio is $R=D/d_f=7$, a fine particle is never trapped in a bed formed by monodisperse larger particles (diameter $D$). However, when the bed particles have a size distribution half-width of $w=0.1$, the same fine particle can be trapped, since the effective trapping threshold is then $R_t=7.182$. 

\subsection{Simulation methodology}

DEM simulations are used to explore the effects of size-polydispersity on the percolation of fine particles. We first generate spherical bed particles with mean particle diameter $D=5$ mm. These particles are arranged in a simulation domain that is periodic in the $x$ and $y$ (horizontal) directions to eliminate wall effects, in particular, void spaces between particles and a rigid wall that are larger than the spaces between particles \cite{leeMicaceousSandsMicroscale2007,suzukiStudyWallEffect2008}. 
The horizontal dimensions of the domain are both $20D$. For polydisperse beds, the bed particle diameter has a volume-based uniform distribution between $(1-w)D$ to $(1+w)D$, considering cases for $w$ in increments of 0.1 from 0 to 0.5. 
To generate smooth particle size distributions, the diameter increment for the discrete particle diameters is set to 0.05 mm. For example, there are 21 distinct diameters ranging from 4.5 mm to 5.5 mm for $w=0.1$, and 101 distinct diameters ranging from 2.5 mm to 7.5 mm for $w=0.5$. The total volume (or mass since particle density is fixed at 2500\,kg\,m$^{-3}$) of particles for each diameter increment is identical. Consequently, the number of bed particles for each particle diameter increment is different, decreasing as species size increases. For the case with the largest diameter particles ($w=0.5$), only 0.87\% of bed particles have a diameter of 7.5 mm but 19.89\% have a diameter of 2.5 mm. All other cases (with smaller $w$) have a larger number of largest particles and a smaller number of smallest particles. A similar approach is used for bidisperse beds, except only two sizes of particles are used, $(1-w)D$ and $(1+w)D$.

Bed particles generated according to the volume-based distribution are released from their initial positions on a cubic lattice and settle in the periodic box under gravity $g=9.81$ m\,s$^{-2}$ (in the negative $z$-direction). Since our focus here is on understanding the effects of polydispersity in isolation without the added complications introduced by bed particle overlaps, a stiff linear spring ($k_n = 10^{12}$) is used in the DEM contact model to ensure that bed particles overlap by less than 0.1\% and to mitigate the effect of the hydrostatic pressure gradient on particle overlap deep in the bed. The bed particles are then frozen in place, after ensuring their positions have remained unchanged for at least 1 s.  Exactly 15000 bed particles fill the box to form the packed bed with bed height ranging from about $25D$ to $30D$. Due to the fixed number of bed particles, the static packing of monodisperse particles has the largest total volume (or height), and the bed volume decreases as the size distribution half-width $w$ increases. Similarly, the packing density increases from about 0.624 for $w=0$ to 0.655 for $w=0.5$ in polydisperse beds and to 0.624 for $w=0.2$ in bidisperse beds.
This phenomenon is consistent with the much more pronounced effect for the packing of size-bidisperse particle mixtures, \cite{petitAdditionalTransitionLine2020} where the packing density can approach 0.85 when the large to fine particle size ratio exceeds five. For this reason and to avoid the free sifting of bed particles themselves, we use $w\leq0.5$, for polydisperse bed particles, to achieve an overall homogeneous packing. For bidisperse bed particles, we maintain $w=0.2$ and the packing density at 0.624, while varying the concentration of small particles.

Single fine particles, with identical diameter, $d_f$, are added above the free surface at random locations within a height difference less than $2D$ and with zero initial velocity, and then free fall under gravity. Similar to our previous work, we simulate single fine particles, i.e., fine particles do not interact with each other (but do interact with bed particles), to avoid the situation where multiple fine particles jam a pore throat \cite{gaoPercolationFineParticle2023a}. The number of fine particles is 5\,000 in order to provide adequate statistics in one simulation. We use the Hertz contact model with a Young's modulus of $5\times10^7$\, Pa, a Poisson's ratio of 0.4, a restitution coefficient of 0.8, and a friction coefficient of 0.5 for both the percolating fine particles and the bed particles, which were previously frozen in place. The simulation time step is $5\times10^{-6}$\,s to assure numerical stability, especially for large size ratios. 

\subsection{Effect of size distribution --- Polydisperse beds}

\begin{figure*}[t]
    \centering
    \includegraphics[width=\linewidth]{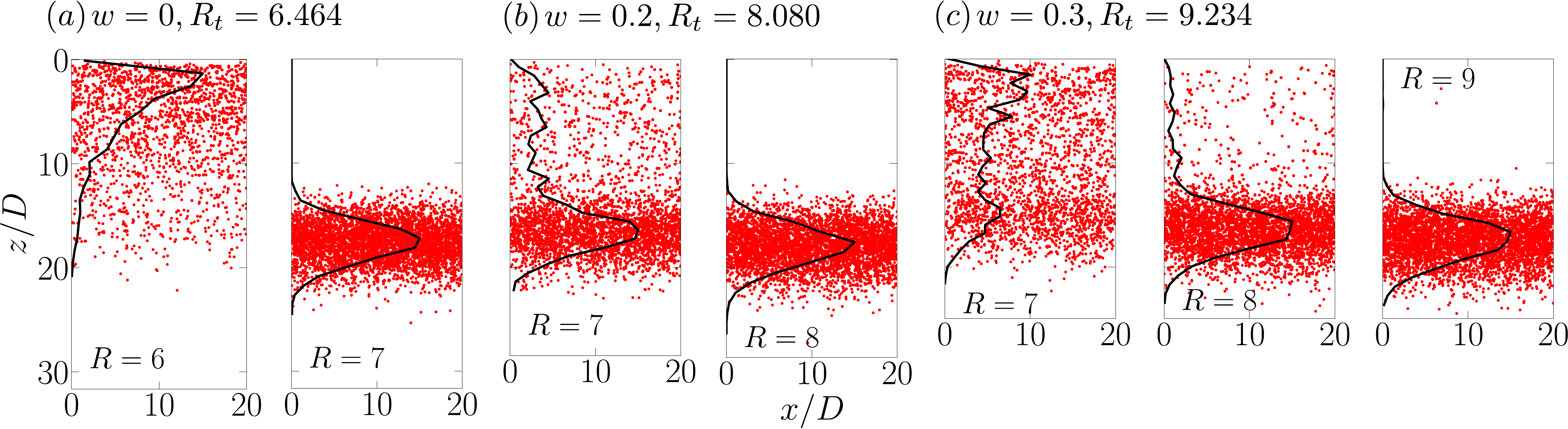}
    \caption{Fine particle percolation for various bed-to-fine particle size ratios, $R=D/d_f$,  in packed beds with different particle size distribution half-width: (a) $w=0$ (monodisperse), (b) $w=0.2$, and (c)  $w=0.3$.  Red dots show the positions of 5000 fine particles at $t=1.5$\;s. Large particles forming the static packed bed are not shown. Black curves are the number spatial distributions of fine particles vs. bed height.}
    \label{fig:chporefig2}
\end{figure*}

To demonstrate the effect of the width of the bed particle size distribution on fine particle percolation, we consider various large to fine particle size ratios, $R=D/d_f$, above and below  $R_t$ values corresponding to different values of $w$, see Fig.~\ref{fig:chporefig2}. All snapshots of the simulations are taken 1.5~s after fine particles are added at the top of the packed bed (large particles forming the packed bed are not shown). 
The top free surfaces for all cases are aligned to allow comparison of the depth that the fine particles (red dots) reach.
It is evident from the maximum vertical depth, $z/D$, that the depth of the bed decreases as $w$ increases due to increased packing density for wide particle size distributions. As shown in Fig.~\ref{fig:chporefig2}, fine particle percolation behavior depends on the particle size ratio $R$ relative to the trapping threshold $R_t$ resulting from the bed particle size distribution half-width, $w$. In the size-monodisperse packed bed ($w=0$), with $R=6$, 97\% of fine particles are trapped after 1.5~s, evident as particles retained near the upper surface, since $R<R_t$ (equivalent to $d_f>d_p$). Of course, all fine particles will eventually be trapped, given longer simulation time. In contrast, fine particles continue percolating without being trapped for $R=7$, evident as a band of particles more than halfway to the bottom of the bed. These two cases correspond exactly to the trapping and passing regimes observed previously \cite{gaoPercolationFineParticle2023a}.  
The fine particle spatial number distribution follows an approximately log-normal relationship for the trapping regime and an approximately normal distribution for the passing regime (black curves), consistent with previous results \cite{gaoPercolationFineParticle2023a}. 

When $w$ is increased to 0.2, the trapping threshold becomes $R_t=8.080$, according to Eq.~\ref{eq:chporeeq1} for a triad of the smallest particles. However, fine particles with $R=8$ (right column of Fig.~\ref{fig:chporefig2}(b)) are seldom trapped, even though $R < R_t$. Although the SSS condition describes the worst case, most bed particle triads in random packings form larger pore throats, such as the LLL, SLL, or SSL conditions. Nevertheless, a significant portion of fine particles, about 39\%, with $R=7$ are trapped at 1.5~s (left column of Fig.~\ref{fig:chporefig2}(b)), even though $R=7$ is well above $R_{t0}=6.464$ based on the nominal bed particle size. 
Of course, all fine particles for the $R=7$ case will be trapped given sufficient simulation time. Nonetheless, Fig.~\ref{fig:chporefig2}(b) indicates that most pore throats are larger than the theoretical smallest bound. 
Moreover, in physical random packings, the bed particle size distribution is more likely to be normal or log-normal rather than the uniform distribution used here, so the probability of the formation of the smallest pore throat may be different from that indicated here. 

For $w=0.3$, where the effective trapping threshold is $R_t=9.234$, very few ($<$0.7\%) fine particles are trapped with $R=9$, but a small number of fine particles (about 19\%) with $R=8$ and a majority of fine particles (about 76\%) with $R=7$ are trapped by the packed bed within 1.5~s of simulation time, as shown in Fig.~\ref{fig:chporefig2}(c). Again, given sufficient simulation time, all fine particles will be eventually trapped for $R=7$ and 8. The key point is that even though the nominal particle size ratio is $R>R_{t0}$ (suggesting that no fine particles would be trapped),  significant numbers of fine particles can become trapped when $w=0.2$ or $0.3$.

\begin{figure}[t]
    \centering
    \includegraphics[width=0.45\linewidth]{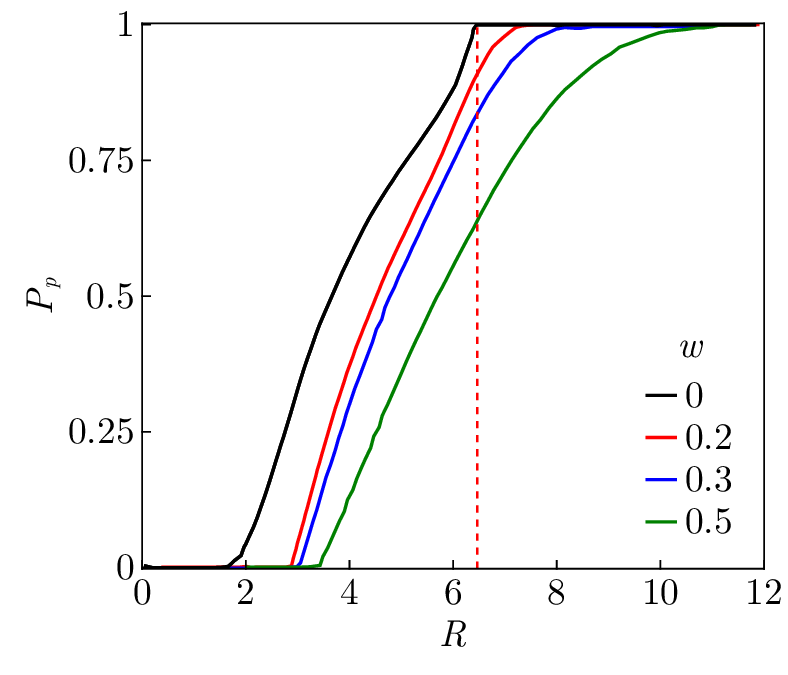}
    \caption{Passing probability $P_p$ of fine particles having size ratio $R=D/d_f$ through static packings formed by size-polydisperse bed particles with different width size-distributions. The red vertical dashed line denotes $R_{t0}=6.464$ for monodisperse ($w=0$) bed particles. }
    \label{fig:chporefig4}
\end{figure}

To quantify the effects of varying bed particle sizes on the pore throat size distribution, we employ Delaunay triangulation and additional analysis to characterize pore sizes \cite{reboulStatisticalAnalysisVoid2008}.
Using this approach Gao et al. \cite{gaoPercolationFineParticle2023a} demonstrate that the cumulative density function (CDF) of pore throat size can be used to determine the passing probability for a fine particle of a specific size, denoted as $P_p = 1 - \textit{CDF(R)}$. This passing probability represents the fraction of pore throat diameters in a large particle packed bed that exceeds the fine particle diameter.
Fig.~\ref{fig:chporefig4} illustrates the passing probability for different half-widths ($w$) of the bed particle size distribution.
As expected for a packed bed comprising identical bed particles ($w=0$), the passing probability ($P_p$) is 1 for $R > 6.464$, but for particles with $R = 6$, the passing probability is 88\%, indicating that 12\% of the pores are smaller than the fine particles, resulting in particle trapping evident on the left side of Fig.~\ref{fig:chporefig2}(a). On the other hand, for $w=0.2$, Fig.~\ref{fig:chporefig4} indicates that the passing probability at $R=7$ is 97\%, corresponding to only 3\% of the pores being smaller than the fine particles.  Hence, a relatively small degree of fine particle trapping occurs for $R=7$ in Fig.~\ref{fig:chporefig2}(b). Fig.~\ref{fig:chporefig4} indicates that for $w = 0.3$, the passing probabilities for $R = 7$ and $8$ are 93\% and 98\%, respectively, resulting in more fine particles being trapped for $R = 7$ than $R = 8$ in Fig.~\ref{fig:chporefig2}(c). For $R = 9$, the passing probability is 99.4\%, so most fine particles percolate freely in Fig.~\ref{fig:chporefig2}(c).
We note in Fig.~\ref{fig:chporefig4} that when $w = 0.2$, $P_p$ reaches unity for $R > 7.8$ a value that deviates from $R_t=8.08$ for $w = 0.2$ calculated from Eq. \ref{eq:chporeeq1}. This discrepancy arises because any given triad of particles may not be in perfect contact in actual particle beds. This variation is also evident in Fig.~\ref{fig:chporefig4} for $w = 0.3$, where $P_p = 1$ is attained at $R > 8.1$, even though $R_t = 9.234$ for $w=0.3$.
Nevertheless, these results demonstrate the sensitivity of fine particle percolation to variations in polydispersity. Therefore, in both experimental and DEM studies focused on particle trapping, it is crucial to account for the effects of bed-particle size-polydispersity.

\subsection{Effect of contact configuration distribution --- Bidisperse beds}

Having examined how bed particle polydispersity influences pore throat sizes, we now investigate the effect of various contact configurations of triads forming the pore throats (LLL, SSS, SSL, and SLL) by implementing a bidisperse bed and varying the relative volumetric concentrations of small and large particles. 
We consider five small particle concentrations, $c_s = $ 0.1, 0.25, 0.5, 0.75 and 0.9, while maintaining the bidispersity at w=0.2, with a packing density of 0.624 and bed dimensions of $20D \times 20D \times 32D$.
Not all triads in a randomly packed bed are expected to be in perfect contact, in contrast to the schematic shown in Fig.~\ref{fig:chporefig1}. Therefore, the configurations are based on the sizes of the particles forming the pore throat and not their state of contact. 

\begin{figure}[h ]
    \centering
    \includegraphics[width=0.5\linewidth]{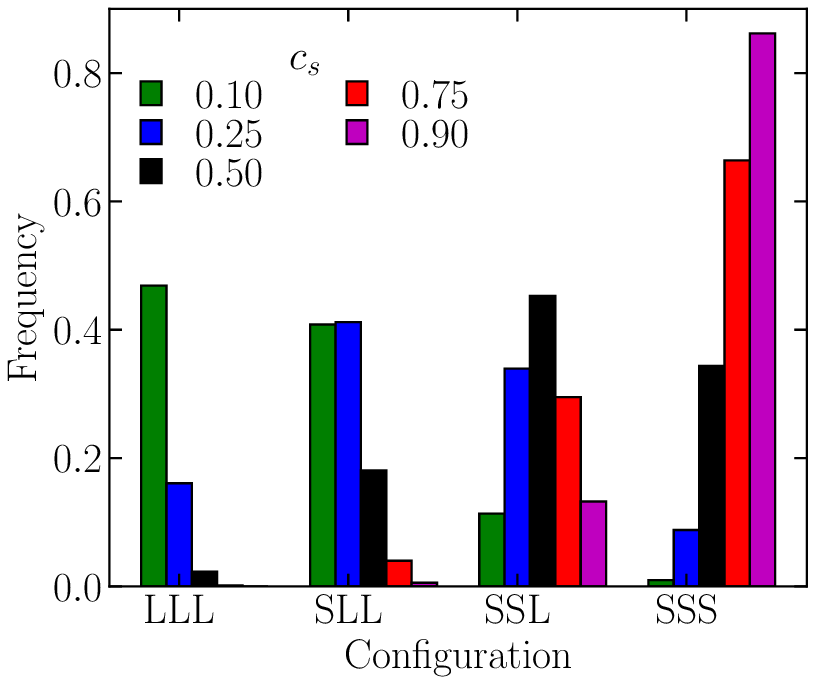}
    \caption{Contact configuration relative frequencies for various small particle concentrations, $c_s$, in a size-bidisperse bed with $w=0.2$.}
    \label{fig:ll}
\end{figure}

The relative frequencies of various contact configurations are shown in Fig. \ref{fig:ll}. For $c_s = 0.1$, the dominant configurations are LLL and SLL. For $c_s = 0.25$, the dominant configurations are SLL and SSL. For $c_s = 0.5$, SSL configurations are the most common, followed by SSS configurations. For $c_s = 0.75$ and  $c_s = 0.9$, the SSS configuration is dominant with the SSL configuration being the second most likely.  

\begin{figure}[h]
    \centering
    \includegraphics[width=0.5\linewidth]{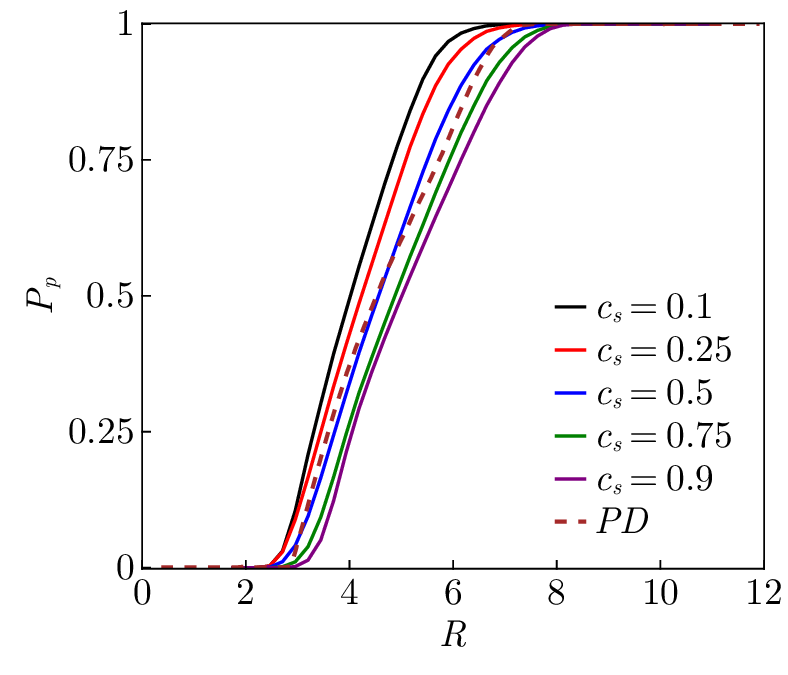}
    \caption{Passing probabilities $P_p$ for different small particle concentrations $c_s$. The dashed curve is the passing probability for the polydisperse case (PD), with $w=0.2$, described in Section 2.2.}
    \label{fig:pp}
\end{figure}

The impact of the resulting pore throat configurations is evident in the passing probabilities shown in Fig. \ref{fig:pp}. 
The highest passing probability at a particular size ratio occurs for $c_s=$ 0.1, where relatively larger configurations (LLL and SLL) are dominant, followed by $c_s =$ 0.25, 0.5 and 0.75. For $c_s = 0.9$, where the SSS configuration is most common, passing probabilities are lowest. 
This demonstrates that not only does the dispersity affect percolation behavior, but the relative concentrations of different bed particle sizes can also play a significant role. 
The passing probability for the polydisperse (PD) configuration shown in Section 2.1 closely matches the $c_s=0.5$ passing probability, indicating that, for this case, the net impact on the passing probabilities is similar when the concentration-weighted average diameter of the bed particle is similar. 

\begin{figure*}[t]
    \centering
    \includegraphics[width=0.8\linewidth]{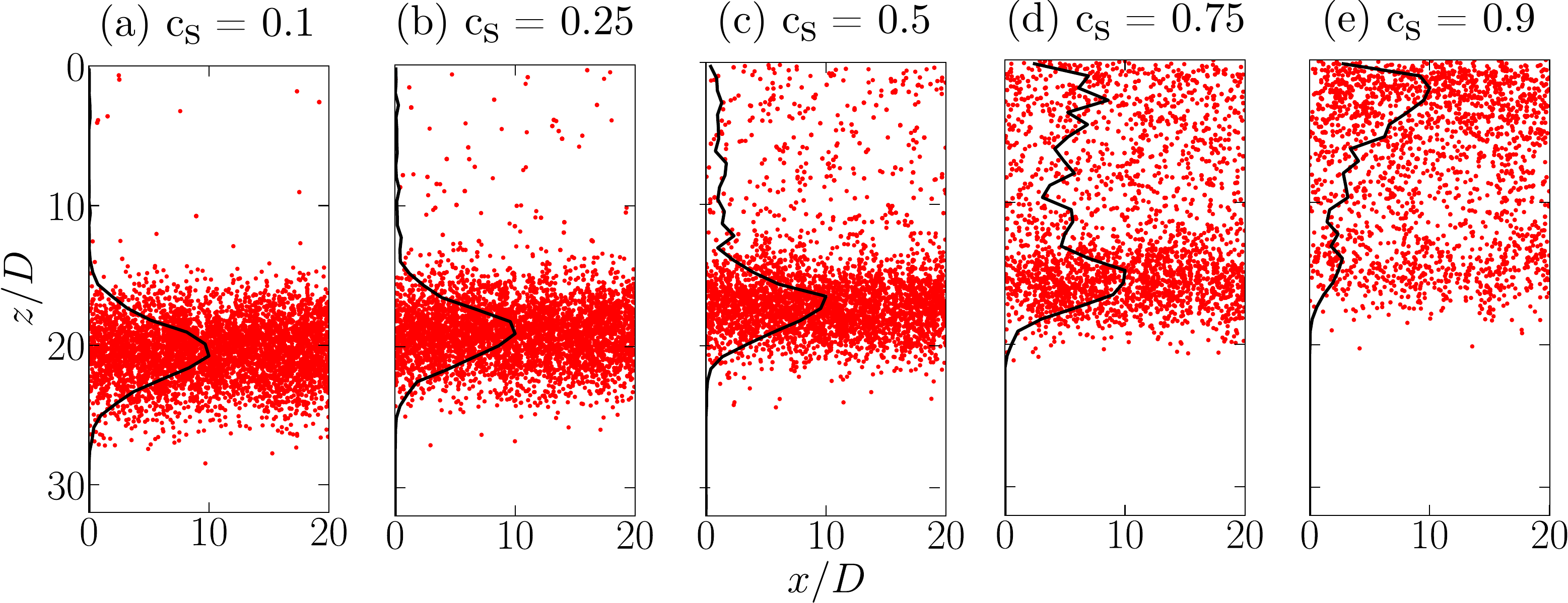}
    \caption{Fine particle percolation in bidisperse beds with $w = 0.2$, and with various small particle concentrations. Red dots show the positions of 5000 fine particles ($R =$ 7) at $t=1.5$\;s. Particles forming the static-packed bed are not shown. Black curves are the number spatial distributions of fine particles vs. bed height.}
    \label{fig:conc}
\end{figure*}

The effect of relative concentration is further elucidated in Fig. \ref{fig:conc}, where the positions of fine particles at the end of 1.5 s are shown for $R = 7$. Consistent with the passing probabilities, the fine particle penetration in the bed is greater at the lower concentrations ($c_s = 0.1$ and $c_s = 0.25$), whereas for $c_s = 0.75$ and $c_s = 0.9$ a greater number of fines become trapped and the penetration depth is reduced due to the prevalence of SSS and SSL contact configurations.

\section{Effect of deformation on trapping}
\label{sec:def}

Particle deformation, which occurs in a variety of physical situations including powder compaction in pharmaceuticals \cite{sunRelationshipTensileStrength2018}, bead deformation in ion-exchange beds \cite{tiihonenElasticityIonexchangeResin2001} and soft microparticles in biomedical applications \cite{dalyHydrogelMicroparticlesBiomedical2020}, can drastically change the pore throat size.  As a case study, here we focus on the influence of particle deformation on pore throats in DEM simulations, which are often used to study particulate beds and granular flows. In DEM simulations, repulsive particle-particle normal forces are represented by particle overlaps \cite{cundallDiscreteNumericalModel1979a}. In simplified models, such as the linear spring model \cite{waltonApplicationMolecularDynamics1984a}, the amount of overlap, which is analogous to the particle deformation for a specific force, is determined by the DEM model spring stiffness, which can be linked to material properties \cite{thorntonInvestigationComparativeBehaviour2011b}. More physically realistic models like the Hertzian model \cite{hertzMiscellaneousPapers1896a,johnsonContactMechanics1985a} establish a non-linear correlation between force and overlap that depends on particle size, Young's modulus, and Poisson ratio.
Such models demonstrate accurate behavior prediction for deformations extending up to 10\% \cite{lommenSpeedupStiffnessEffects2014d,paulickReviewInfluenceElastic2015a, giannisStressBasedMulticontact2021}.




Calculating the trapping threshold for overlapping size-monodisperse particles is straightforward. In the schematic shown in Fig.~\ref{fig:sch}(c), the centers of the three overlapping spheres (with diameter $D$ and overlap $\delta$) create an equilateral triangle, with each edge measuring $D-\delta$ in length. By identifying the center and diameter of a circle that touches all three spheres, we can determine the pore throat diameter ($d_p$) as 
\begin{equation}
    \frac{d_p}{D} = \frac{2-\sqrt{3}}{\sqrt{3}} - \frac{2}{\sqrt{3}}\frac{\delta}{D}.
    \label{eq:delta_dc}
\end{equation}
The trapping threshold, 
\begin{equation}
    R_t = D/d_p = \frac{\sqrt{3}}{(2-\sqrt{3})-2\delta/D}, 
    \label{eq:rtdel}
\end{equation}
is, therefore, inversely related to the overlap ($\delta$).

\begin{figure}[h]
    \centering
    \includegraphics[width=0.6\linewidth]{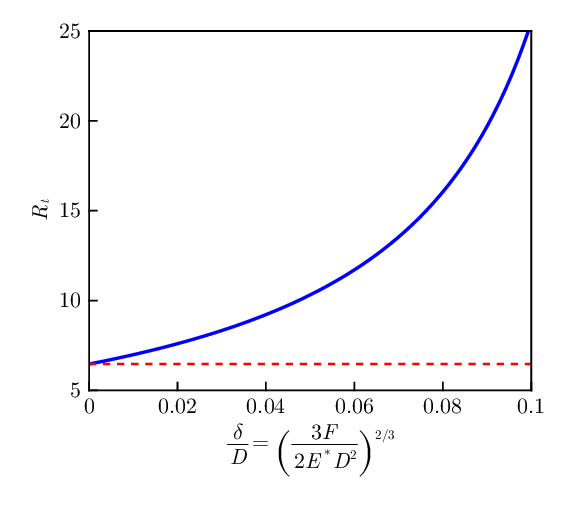}
    \caption{Variation of trapping threshold, $R_t$, with scaled particle overlap $\delta/D$. Red dashed line indicates $R_{t0} = 6.464$ for non-deforming particles.}
    \label{fig:sr_delta}
\end{figure}

In physical systems, the pore throat size may not be specified in terms of overlaps, but rather in terms of material properties and the applied overburden pressure or force that prescribes the particle deformation at the contact point. For the Hertzian contact force model, 
\begin{equation}
    F = \frac{4}{3} E^* r^{*^{1/2}} \delta^{3/2}.
    \label{eq:hertz}
\end{equation}
Here, $F$ represents the pair-wise applied force, $E^*$ is the effective Young's modulus, 
\begin{equation}
    \frac{1}{E^*} = \frac{1-\nu_1^2}{E_1} + \frac{1-\nu_2^2}{E_2},
    \label{eq:estar}
\end{equation}
and $r^*$ is calculated as
\begin{equation}
    \frac{1}{r^*} = \frac{1}{r_1} + \frac{1}{r_2},
    \label{eq:rstar}
\end{equation}
where, the subscripts 1 and 2 represent the two contacting particles, and $E_i$, $\nu_i$, and $r_i$ are their Young's modulus, Poisson ratio, and radius, respectively.  Rewriting Eq. \ref{eq:hertz} in terms of overlap for the identical mono-disperse large particles with $D$ and $E^*=E/(1-\nu^2)$ gives
\begin{equation}
    \frac{\delta}{D} = \left(\frac{3}{2}\frac{F}{E^*D^2}\right)^{2/3}.
    \label{eq:delhertz}
\end{equation}
The dependence of the trapping threshold $R_t$ on the nondimensional overlap $\delta/D$ (Eq. \ref{eq:rtdel}) or, alternatively, the nondimensional contact force $F/E^*D^2$ (Eq. \ref{eq:delhertz}), is shown in Fig.~\ref{fig:sr_delta}.

The deformation, or overlap $\delta$ in DEM simulations, is zero when either the applied force $F = 0$ or the material is rigid $E^*\to\infty$. For these cases the trapping threshold in Fig.~\ref{fig:sr_delta}  occurs at $R_t = R_{t0} = 6.464$.
As the overlap increases, the pore throat diameter ($d_p$) decreases, leading to a reduction in the size of fine particles that can pass through the pore throat. As indicated in Fig.~\ref{fig:sr_delta}, even for relatively small overlaps, there is a significant increase in $R_t$. For instance, with $\delta/D=0.02$, the trapping threshold rises to $R_t = 7.6$.
In Eq. \ref{eq:rtdel}, $R_t$ diverges to $\infty$ at $\delta/D = (2-\sqrt{3})/2 = 0.134$. 
Beyond this overlap value, the pore throats are sealed, preventing any fine particles from passing through. In the context of observed overlaps in DEM simulations ranging from 0\% to 10\% \cite{lommenSpeedupStiffnessEffects2014d,paulickReviewInfluenceElastic2015a, giannisStressBasedMulticontact2021}, it is evident that even a relatively small overlap of 1\%, which increases $R_t$ to nearly 7, may need to be taken into account, depending on the application.

\subsection{Effect of material properties}

Granular packed beds are used across diverse applications, with the Young's moduli of the particles ranging from a few kilopascals (e.g. 20 kPa for hydrogels \cite{dalyHydrogelMicroparticlesBiomedical2020} employed in biomedical applications) up to several gigapascals (e.g. 200 GPa for steel \cite{nayarMetalsDatabook1997}). The Poisson ratios for these materials vary between 0.2 and 0.4. Eqs. \ref{eq:rtdel} and \ref{eq:delhertz} provide the relationship between material properties ($E$ and $\nu$), the applied force ($F$), particle deformation ($\delta$), and the trapping threshold ($R_t$).

\begin{figure}[t]
    \centering
    \includegraphics[width=0.6\linewidth]{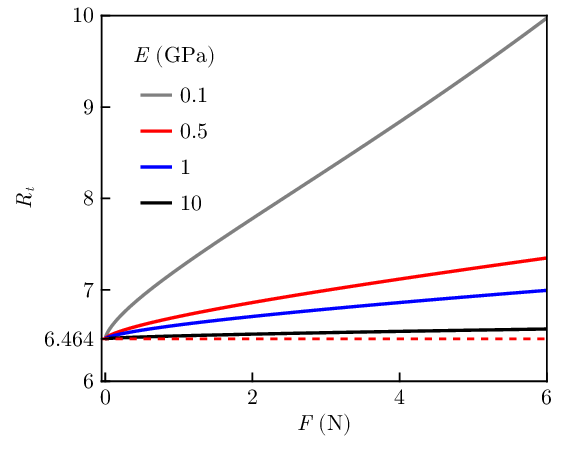}

    \caption{Variation of the trapping threshold, $R_t$, with contact force, $F$, for four values of Young's modulus, $E$.}
    \label{fig:sr_f}
\end{figure}

As examples, consider four Young's modulus values ($E = 0.1$ GPa, $0.5$ GPa, $1$ GPa, and $10$ GPa) and a 4 mm diameter bed particle with a Poisson ratio of $\nu = 0.3$. 
As shown in Fig. \ref{fig:sr_f}, for a given applied force $F$, softer materials (lower $E$), exhibit larger overlaps, leading to higher trapping thresholds, while the trapping threshold is barely affected for large $E$. For instance, for $E=10$ GPa, at $F$ = 6 N, the trapping threshold is $R_t = 6.57$, but increases to $R_t = 9.84$ for $E=0.1$ GPa.

\subsection{Simulation methodology}

\begin{figure}[t]
    \centering
    \includegraphics[width=0.3\linewidth]{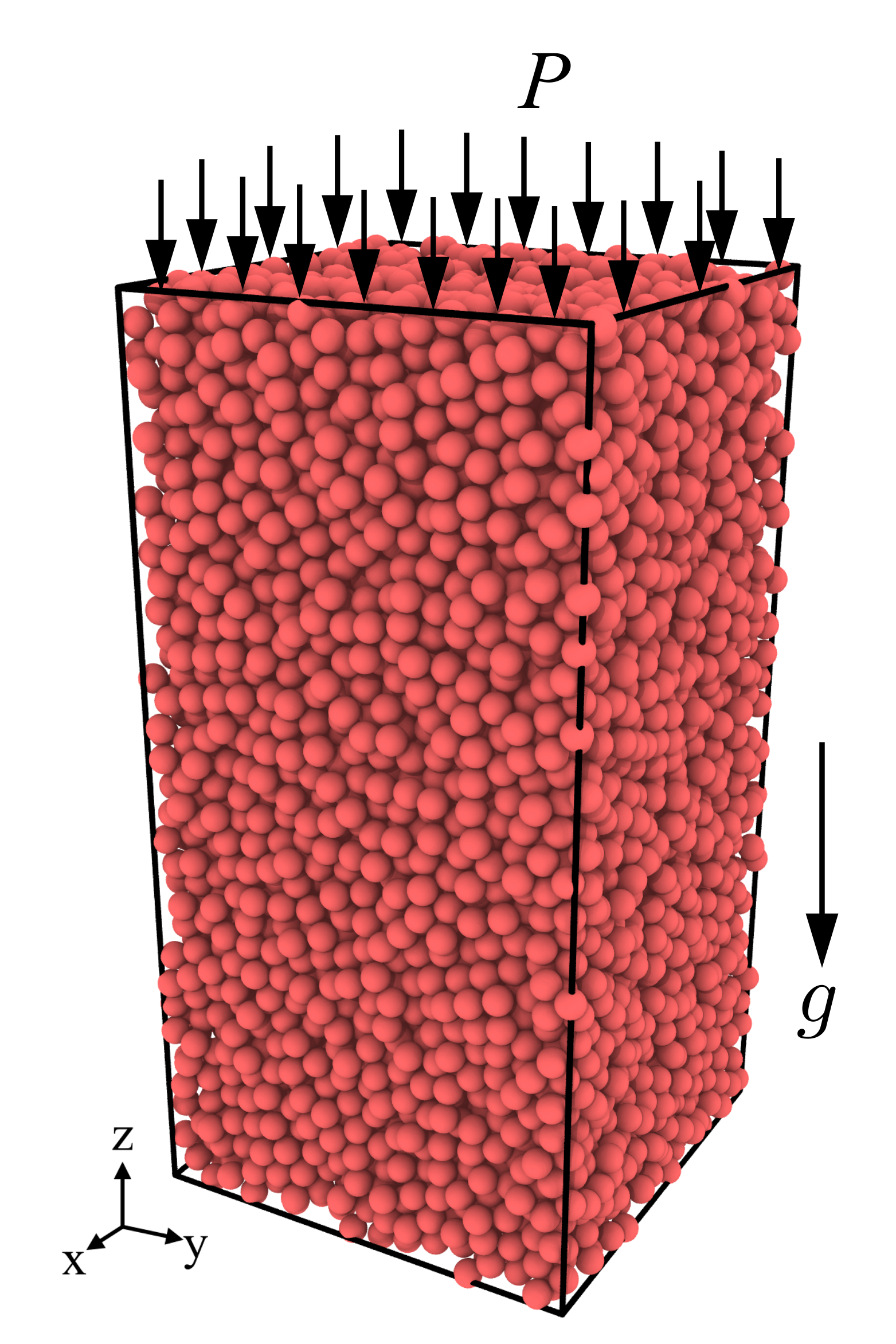}
    \caption{Static random bed of large monodisperse particles with an overburden pressure $P$. Fine particles (not shown) percolate downward through the bed under the influence of gravity.}
    \label{fig:dem}
\end{figure}

To demonstrate the influence of this variability of particle deformation in practical applications, we analyze a randomly packed bed with an overburden pressure $P$, as illustrated in Fig.~\ref{fig:dem}. The bed is composed of $10^4$ uniformly sized large particles having diameter $D = 4$ mm, density $\rho = 2500$ kg/m$^3$, Poisson ratio $v = 0.3$, restitution coefficient of 0.8, and friction coefficient of 0.5. 
The bed is periodic in the horizontal directions ($x$ and $y$) and has sides of $16D$. The random packed bed is prepared using the particle growth algorithm proposed by Lubachevsky and Stillinger \cite{lubachevsky_geometric_1990} and used in \cite{gaoPercolationFineParticle2023a}. 
Four different values of Young's modulus are considered: $E = 0.1$ GPa, $0.5$ GPa, $1$ GPa, and $10$ GPa. 
In order to analyze the impact of the applied force, two values of the applied overburden pressure, $P = 10^3$ Pa  ($\approx$ 0.01 atm) and $10^4$ Pa ($\approx$ 0.1 atm)  are considered, which for the chosen density and packing fraction correspond to lithostatic pressures for bed heights of 0.068 m and 0.68 m, respectively. Gravity does not act on the bed particles to assure a homogeneous distribution of applied forces from top to bottom of the bed.

\subsection{Overlap effects in randomly packed beds}

\begin{figure*}[t]
    \centering
        \centering
        \includegraphics[width=0.9\linewidth]{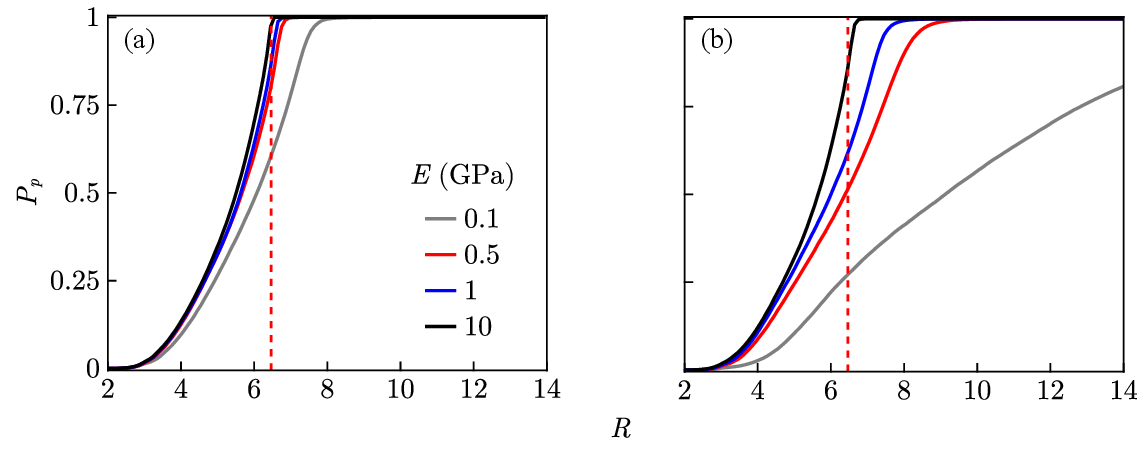}
    \caption{Passing probability, $P_p$, of the pore throat size for random packings with different values of Young's modulus and overburden pressure: (a) $P = 10^3$ Pa and (b) $P = 10^4$ Pa. Red dashed line indicates $R_{t0} = 6.464$.}
    \label{fig:cdf}
\end{figure*}

\begin{figure*}[t]
    \centering
        \includegraphics[width=0.9\linewidth]{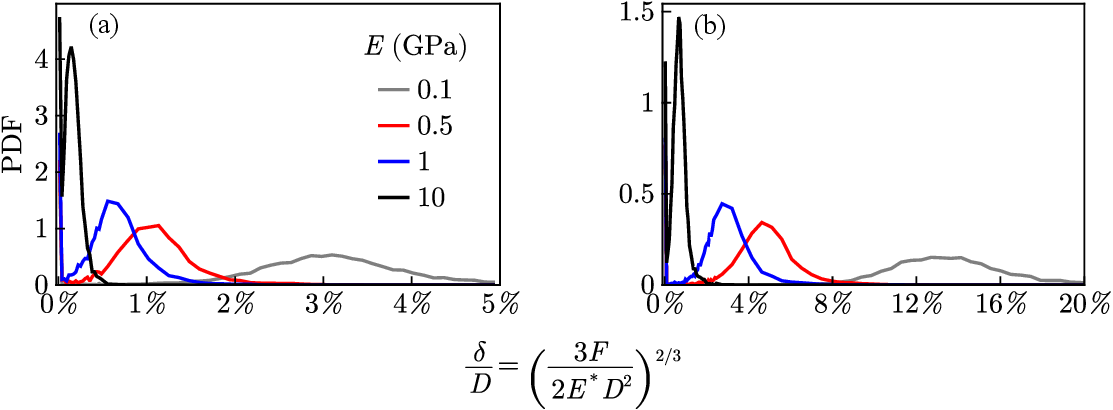}
    \caption{Probability density function (PDF) of particle overlaps in the packed beds for overburden pressures of (a) $10^3$ Pa and (b) $10^4$ Pa.}
    \label{fig:ovl}
\end{figure*}

To understand the effect of bed particle overlap on fine particle percolation, we analyze the passing probabilities $P_p(R)$, for various values of $E$ under two applied pressure conditions, as shown in Fig.~\ref{fig:cdf}. For an applied pressure of $10^3$ Pa, the passing probabilities for materials with $E\ge0.5$ GPa exhibit minimal sensitivity to Young's modulus ($E$). Here, $P_p$ reaches a value of one within the range of $6.5<R<7$, indicating that fine particles with $R>7$ can freely percolate (that is, $d_f < d_p$). However, for $E = 0.1$ GPa, the passing probability reaches unity only when $R = 8$.
This effect is also evident in the distribution of particle overlaps measured in the granular beds, shown in Fig.~\ref{fig:ovl}(a). 
The distribution of overlaps indicates that different pairs of particles in different triads have different amounts of deformation, thereby influencing the local behavior of fine particles traversing the bed. 
For $E = 0.1$ GPa, the mean dimensionless overlap is $\delta/D \approx $ 3\%, which decreases to about 1\%, 0.65\%, and 0.15\% for $E = 0.5$ GPa, $1$ GPa, and $10$ GPa, respectively, with corresponding values for $R_t$ of 8.34, 7.02, 6.79, and 6.54, respectively, as determined from Eq. \ref{eq:rtdel}. These values align approximately with the values for $R$ at which $P_p$ reaches one in Fig.~\ref{fig:cdf}(a).
Hence, the mean values of the distributions in Fig.~\ref{fig:ovl}(a) correspond approximately to deformations for the size ratio at which free sifting occurs.
Furthermore, while Eq. \ref{eq:delhertz} provides insights into how the trapping threshold varies with overlap and applied force, the forces and overlaps in actual particle beds are typically not uniform in magnitude \cite{jaegerGranularSolidsLiquids1996}. 

In Fig.~\ref{fig:cdf}(b), the influence of $E$ becomes more pronounced as the pressure is increased to $P = 10^4$ Pa. For $E = 10$ GPa, the size ratio $R$ at which $P_p$ reaches one is 6.8, which increases to 8.6 and 10 for $E = 1$ GPa and $E = 0.5$ GPa, respectively. However, for $E = 0.1$ GPa, the passing probability in Fig.~\ref{fig:cdf}(b) never reaches one before $\delta/D$ exceeds the critical value of 13.4\%, which corresponds to pore sealing.
The distribution of overlaps for this case in  Fig.~\ref{fig:ovl}(b) indeed indicates that the overlap exceeds 13.4\% for about half of the overlaps, consistent with pore sealing.  On the other hand, for $E =$ 10, 1, and 0.5 GPa, the mean values of $\delta/D$ are approximately 0.7\%, 3.0\%, and 4.7\%, respectively, with corresponding $R_t$ values of 6.80, 8.29, and 9.91. Again, these values are close to the values at which $P_p$ reaches one in Fig.~\ref{fig:cdf}(b), which are 6.8, 8.6 and 10, respectively.
This underscores the feasibility of estimating the size ratio at which fine particles can percolate through a bed of deformed large particles based on the average particle overlap.

\begin{figure*}[t]
    \centering
        \centering
        \includegraphics[width=0.75\linewidth]{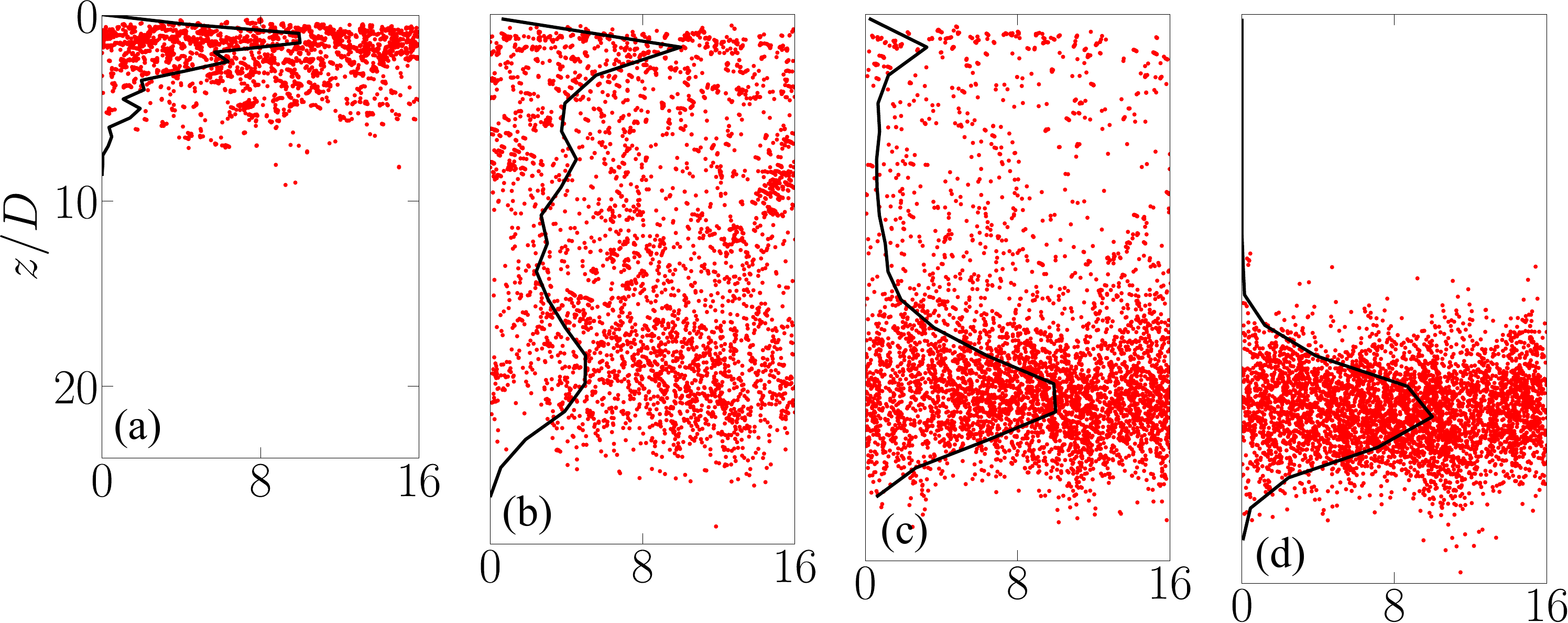}
    \caption{Fine particle percolation in packed beds with different Young's moduli:  (a) $E = 0.1$ GPa, (b) $E = 0.5$ GPa, (c) $E = 1$ GPa and (d) $E = 10$ GPa and an overburden pressure of $P = 10^4$ Pa.
    Red dots show the positions of 5000 non-interacting particles having size ratio $R = 8$, at $t=1.25$ s. Large particles forming the packed bed are not shown. Black curves are the number spatial distributions of fine particles vs. bed height.}
    \label{fig:E}
\end{figure*}

The interplay between the Young's modulus, overlap, and applied pressure plays a crucial role in determining the trapping behavior of granular systems. 
To underscore this point, Fig.~\ref{fig:E} presents snapshots (at $t$ = 1.25 s) of 5000 fine particles with a size ratio $R = 8$ percolating through a bed of large particles under the influence of gravity ($g = 9.81$~m/s$^2$) for different values of $E$ and an overburden pressure of $P = 10^4$ Pa.  
The fine particles interact with the bed particles but not each other. Although the beds contain the same number of large particles of identical size, the applied pressure compresses the beds composed of softer material more than those made of stiffer material.
%
 For $E = 0.1$ GPa, the passing probability for $R = 8$ is 43\% (see Fig.~\ref{fig:cdf}(b)), signifying that 57\% of pore throats within the bed are smaller than the particle size. Consequently, all of the fine particles are trapped near the top of the bed.
For $E = 0.5$ GPa, the passing probability is 78\% resulting in  56\% of the fine particles still percolating at $t=$ 1.25 s, while the remaining are trapped within the bed (and all particles will eventually be trapped at longer times). 
For $E = 1$ GPa, only 7\% of the fine particles become trapped at $t=$ 1.25 s, because $P_p =$ 97\%, allowing for more extensive percolation.
Lastly, for $E = 10$ GPa, $P_p = 1$ at $R = 8$, and all fine particles percolate freely through the bed.
%

\begin{figure*}[t]
    \centering
        \includegraphics[width=0.9\linewidth]{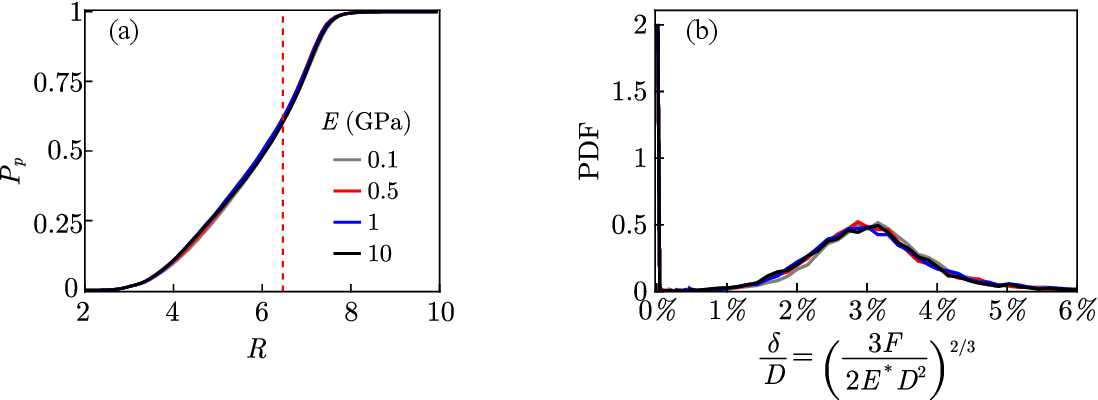}
    \caption{(a) Passing probability, $P_p$, vs. size ratio, $R$, and (b) probability density function (PDF) of particle overlaps, $\delta/D$, in packed beds for different values of Young's modulus $E$ and with an overburden pressure $P = E/10^5$.} 
    \label{fig:ebyp}
\end{figure*}

While the previous examples pertain to specific values of Young's modulus and overburden pressure, the key parameter is the ratio $E/P$. To illustrate this point, we compare four values of Young's modulus when an overburden pressure of $P = E/10^5$ is applied.  The passing probability and the distribution of overlaps shown in Fig.~\ref{fig:ebyp} are almost identical across various cases characterized by the same ratio $E/P$. Minor discrepancies in the overlap distribution can be attributed to random variations in particle positions. This underscores the significance of the $E/P$ ratio as the primary determinant of particle overlap and the resulting fine particle percolation behavior.

\section{Summary}

This study underscores the significance of particle polydispersity and deformation on pore throat size in randomly packed spherical particle beds, which will affect fine particle percolation or fluid flow through packed beds. Figs.~\ref{fig:chporefig1} and \ref{fig:sr_delta} show that the trapping threshold is surprisingly sensitive to both of these variables. 
The implications of these findings are important for practical industrial applications. 
For example, in the design of packed beds of spherical particles for filtration purposes, appropriate selection of the bed particle size distribution and material properties ensures an effective capture or passage of particles within the desired size ranges. Similar effects also need to be considered for granular beds consisting of non-spherical particles, and this can be achieved using the approach outlined in this study. However, in such cases, the particle non-sphericity and orientation will also influence percolation. 
%
This study additionally serves as a caution for research endeavors that examine the percolation of fine particles or flow of fluids through packed particle beds to incorporate considerations of variations arising from polydispersity and deformation, as they can markedly influence the outcomes.
To this end, we list a set of heuristics to highlight key considerations for physical systems and DEM simulations.

Guidelines for physical systems
    \begin{itemize}
        \item Packed beds of polydisperse or bidisperse particles can exhibit a wide range of pore throat sizes. This wide range of sizes can be useful if the goal is to trap a wide range of particle sizes \cite{kerimovMechanicalTrappingParticles2018,bonarensPolydispersePackedBed2022}, but if the goal is to control the size of particles to be trapped, a large range of bed particle sizes can be problematic. 
        \item Packed beds of stiff particles undergo little deformation under pressure and, hence, the pore throat size changes minimally. In contrast, packed beds composed of soft particles deform substantially, even at low pressures, which can significantly reduce the minimum throat size. Therefore, when designing packed particle beds, appropriate consideration of both particle material and overburden pressure is vital.
    \end{itemize}

Guidelines for DEM simulations

\begin{itemize}
    \item In DEM simulations, it is common to introduce a size variation to prevent ordered packing. However, this variation in bed particle size raises $R_t$ significantly.  For instance, a size variation of $\pm$10\% increases the trapping threshold from $R_{t0} = 6.464$ to $R_t \approx 7.2$. Therefore, it is important to exercise caution to steer clear of both ordered packing and particle size variation that could restrict percolation through the bed, especially if this specific scenario is under investigation.
    \item A lower value of Young's modulus in the Hertzian contact model or a lower value of spring stiffness in the linear spring model, while permitting larger time steps in DEM simulations, can result in unreasonably high particle overlaps. This can be a problem for accurately modeling a packed bed. Limiting the overlap to 1\% or less assures that $R_t$ does not exceed 7.
\end{itemize}

While we illustrate the effects of polydispersity and deformation through the example of fine particle free sifting and express the results in terms of the trapping threshold, these results are of broader relevance. For instance, in applications such as oil recovery or groundwater percolation, where fluid flow through pore throats plays a pivotal role, variations in pore throat size can directly impact pressure and flow dynamics. 
In our example of fine particle percolation, even a relatively modest 20 to 30\% bed particle polydispersivity for a packed bed can trap a significant fraction of fine particles that would otherwise percolate through a monodisperse bed of large particles. Likewise, the Young's modulus of the bed particles strongly affects particle deformation, exerting a substantial impact on the percolation behavior of fine particles.

These findings highlight the critical importance of considering the effects of deformation on trapping thresholds.
Understanding how material properties and deformation influence the percolation behavior of particles through constrictions can be leveraged to enhance or reduce solid particle trapping efficiency or fluid flow and optimize related processes.

\section*{Acknowledgments}
    This material is based upon work supported by the National Science Foundation under Grant No. CBET2203703.

\bibliography{refs}
\end{document}